\documentclass[aps,prd,groupedaddress,showkeys]{revtex4-2}
\pdfoutput=1

\usepackage{hyperref}
\usepackage[pdftex]{graphicx}	
\usepackage[version=4]{mhchem}
\usepackage{booktabs}
\usepackage{multirow}
\usepackage{tabularx,booktabs} 
\newcolumntype{Y}{>{\centering\arraybackslash}X} 
\usepackage{url}
\usepackage{soul} 
\begin{document}

\title{Pair production of magnetic monopoles and stable high-electric-charge objects in proton-proton and heavy-ion collisions}

\author{Wen-Yi Song}
\email{wsong@yorku.ca}
\affiliation{Department of Physics and Astronomy, York University, Toronto, ON M3J 1P3 Canada}

\author{Wendy Taylor}
\email{taylorw@yorku.ca}
\affiliation{Department of Physics and Astronomy, York University, Toronto, ON M3J 1P3 Canada}

\date{\today}

\begin{abstract}
We describe pair-production models of spin-0 and
spin-\textonehalf\ magnetic monopoles and high-electric-charge objects
in proton-proton and heavy-ion collisions, considering both the
Drell-Yan and the photon-fusion processes. In particular, we extend the
Drell-Yan production model of spin-\textonehalf\ high-electric-charge
objects to include $Z^0$-boson exchange for proton-proton
collisions. Furthermore, we explore spin-\textonehalf\ and, for the
first time, spin-0 production in ultraperipheral heavy-ion
collisions. With matrix element calculations and equivalent photon
fluxes implemented in \textsc{MadGraph}5\_aMC@NLO, we present
leading-order production cross sections of these mechanisms in
$\sqrt{s} = 14$~TeV proton-proton collisions and $\sqrt{s_{\text{NN}}}
= 5.5$~TeV ultraperipheral lead-lead collisions at the LHC.  While the
mass range accessible in ultraperipheral lead-lead collisions is much
lower than that in proton-proton collisions, we find that the
theoretical production cross sections are significantly enhanced in
the former for masses below 82~GeV.
\end{abstract}

\keywords{magnetic monopoles, high-electric-charge objects, pair production, heavy-ion collisions}
\maketitle

\section{\label{sec:intro}Introduction}

Maxwell's equations possess an electric-magnetic dual symmetry that is
broken by the absence of magnetic charges. In 1931, Dirac showed that
the existence of particles with magnetic charge was consistent with
quantum mechanics~\cite{Dirac:1931kp}. Since then, theories and models
have been developed with different magnetic monopole solutions, as reviewed 
in~\cite{Preskill1984,Shnir,Milton2006,Weinberg2007,Rajantie2012,Mavromatos2020}. Unlike the composite monopoles in the 't Hooft-Polyakov model of grand unification 
theories~\cite{HOOFT1974276,Polyakov:1974ek} or
some extensions of the standard model~\cite{Kephart2017,CHO1997360,Kimm_2015,Ellis2016,Arunasalam2017,Ellis2017,Arai2018,Mavromatos2017,Mavromatos2018,Hung2020,Ellis2020},
Dirac monopoles are point-like and structureless with unknown mass and
spin. Interestingly, the observed quantization of electric charge is
explained by Dirac's quantization condition~\cite{Dirac1948}, which
dictates that a Dirac magnetic monopole carry a magnetic charge that
is an integer multiple of the fundamental Dirac charge $g_\textrm{D}$:
\begin{equation}
\label{eq:ChQuantization}
    \frac{g_\textrm{D} e}{\hbar c} = \frac{1}{2} \longrightarrow
    \frac{g_\textrm{D}}{e} = \frac{1}{2\alpha} \approx 68.5,
\end{equation}
in cgs Gaussian units, where $e$ is the elementary electric charge and
$\alpha \approx 1/137$ is the fine-structure constant.
This implies that the energy loss, or stopping power, of a
Dirac monopole of magnetic charge $1g_\textrm{D}$ in matter is
similar to that of an ion with electric charge $|z|=68.5$,
where $z$ is in units of $e$~\cite{Ahlen1978,Ahlen1980,Ahlen1982}. 
Therefore, magnetic monopoles, as with
high-electric-charge objects (HECOs), are highly ionising particles
(HIPs). Another consequence of the Dirac quantization condition, which
implies that electric and magnetic couplings cannot be simultaneously
small due to their inverse correlation, is that monopoles cannot be
handled with perturbation theory.

Monopole searches at colliders usually target stable Dirac magnetic
monopoles and assume that magnetic charge is 
conserved~\cite{Giacomelli2003,Milton2006,Patrizii2015,Mavromatos2020}. 
Although the 't Hooft-Polyakov monopoles are too massive to be produced
in colliders, numerous attempts have been undertaken to predict monopoles
with masses accessible at colliders such as the 
LHC~\cite{Kephart2017,CHO1997360,Kimm_2015,Ellis2016,Arunasalam2017,Ellis2017,Arai2018,Mavromatos2017,Mavromatos2018,Hung2020,Ellis2020}. 
There is no well-established theory of monopole production to date for direct
cross section calculation due to their strong magnetic
couplings. However, in order to obtain a mass limit that can be
compared to existing experimental results, it is necessary to assume
at least one model with an associated production cross section. Hence,
the common approach is to formulate a magnetic-dual theory of Quantum
Electrodynamics (QED) and consider possible benchmark scenarios. The
benchmark model for monopole production has been the Drell-Yan (DY)
mechanism but the photon-fusion (PF)
mechanism~\cite{Kurochkin2006,2009PhotonFusion,Epele2012,Baines2018} is now of interest, 
since at the current LHC energy scale its cross section in proton-proton
collisions is larger. Contrary to the inelastic Drell-Yan mechanism,
photon fusion encompasses elastic and semi-elastic processes in
addition to inelastic interactions.

To date, monopole and HECO searches at the LHC have been primarily
conducted with proton-proton collisions with the general-purpose detector 
ATLAS as well as the dedicated MoEDAL experiment.
The ATLAS collaboration published the first LHC monopole search in 2012 based on 
7~TeV collision data collected in Run~1~\cite{ATLAS2012} 
and has continuously proven to be particularly sensitive to 
monopoles of charge $1g_\textrm{D}$ and $2g_\textrm{D}$ and HECOs up to $|z|=100$~\cite{ATLAS2013a,ATLAS2016a,Aad_2020}. 
The MoEDAL collaboration released its first monopole search result in 2016
using 8~TeV data from Run 1~\cite{MoEDAL2016a} and has been known for its
capability to detect higher-charge monopoles and 
dyons~\cite{MoEDAL2017a,MoEDAL2018,MoEDAL2019,dyonMoEDAL2021}. 

The LHC not only collides protons but also heavy
ions, which are of great interest due to the large electromagnetic fields
associated with the high-charge ions moving at high velocities.
Previously, monopole searches in heavy-ion collisions were conducted
at the Brookhaven Alternating Gradient Synchrotron (AGS) using gold
nuclei and at the CERN Super Proton Synchrotron (SPS) using lead nuclei
with lead targets, placing lower mass limits of 3.3~GeV and 8.1~GeV,
respectively~\cite{PhysRevLett.79.3134}.
Ultraperipheral collisions, where the ion-ion impact
parameter exceeds the nucleus diameter, could produce magnetic
monopoles via the photon-fusion mechanism. Furthermore,
ultraperipheral collisions have a characteristic signature with low
particle multiplicities, making them easy to select and analyze.

In this paper, we compare the different HIP pair-production mechanisms
in proton-proton and heavy-ion collisions, with an emphasis on
ultraperipheral lead-lead collisions.
Section~\ref{sec:production} identifies the interaction vertices for
HIP production in Drell-Yan and photon-fusion
mechanisms. Section~\ref{sec:heavyIon} introduces heavy-ion collisions
and explains ultraperipheral collisions as one collision type in
lead-lead collisions. Section~\ref{sec:collisions} compares the
parameters of collisions of various heavy ions to those of protons.
Section~\ref{sec:computation} covers the implementation of the
interaction vertices and the equivalent photon
fluxes. Section~\ref{sec:results} presents the cross section results and 
representative kinematic distributions.

\section{\label{sec:production}Partonic mechanisms for HIP pair production}

The approximate duality between electric and magnetic charges in
electromagnetism is the key to the development of a magnetic-dual
theory of QED, which is mediated by photon exchange. QED has an
underlying invariance under $U(1)$ gauge transformations, where the
field of each charged particle picks up a phase proportional to its
electric charge. This invariance fixes the coupling of the field to
the photon and demands electric charge conservation at each
vertex. The electromagnetic coupling to the photon for any
electrically charged particle scales as the charge $|z|$.
Hence, a minimal model of monopole interactions assumes an
electromagnetic monopole-photon coupling that depends on the magnetic
charge $Ng_\textrm{D}$ of the monopole, where $N$ is an integer.
In the case of HECOs, it is completely natural to build additional 
interactions with a gauge invariance under $SU(2)$ transformations 
that govern the Standard-Model weak interactions, where the new 
gauge group gives rise to different matter-gauge couplings.

By constraining the possible interactions, the gauge symmetry allows
us to write candidate Lagrangians from which we can compute the
dynamics of the theory. We assume two spin states for the HIPs: spin-0 and spin-\textonehalf. For a given spin, the candidate
$U(1)$ Lagrangian is the same for monopoles and
HECOs except for the coupling, which depends on
$Ng_\textrm{D}$ for monopoles and $|z|$ for HECOs. Hence, in the
development below, we generalize monopoles and HECOs as HIPs with
charge $g$. Working at leading order, we present two partonic HIP
production mechanisms in proton-proton and heavy-ion collisions. The
Drell-Yan mechanism is the process of a quark and an anti-quark
annihilating to form a virtual photon (or a $Z^0$ boson in the case of HECOs), which then
decays to a pair of HIPs. The photon-fusion mechanism is when two
photons radiated from two colliding hadrons fuse to produce a pair of
HIPs.

Invoking electric-magnetic duality, the charge and coupling of the magnetic
monopole are often assumed to be velocity-dependent. This
choice was based on the observed equivalence, in the small scattering angle
limit, of the electron-monopole scattering differential cross section
in \cite{Schwinger1976} and the Rutherford scattering
differential cross section, after substituting $\frac{g}{c}$ for
$\frac{e}{v}$~\cite{Milton2006}.
The resulting magnetic charge $g\frac{v}{c} \equiv g\beta$,
not only appears in the description of monopole energy loss,
but also when considering monopole pair production via the Drell-Yan
and photon-fusion
mechanisms~\cite{Kurochkin2006,2009PhotonFusion,Epele2012,MoEDAL2016a,MoEDAL2017a,Reis_2017,MoEDAL2018,Baines2018,MoEDAL2019}.
We choose to follow the approach of a minimal
model for Dirac monopoles 
where all HIP charges and couplings are assumed to be velocity-independent.
Doing so leads to slightly different production cross sections
and kinematic features. While the velocity dependence suppresses production of
monopoles with lower velocities, a velocity-independent monopole
coupling does not, thereby giving rise to softer
kinematic distributions and higher production cross sections.

\subsection{Spin-0 interaction}
\begin{figure}[!h]
\centering
\includegraphics{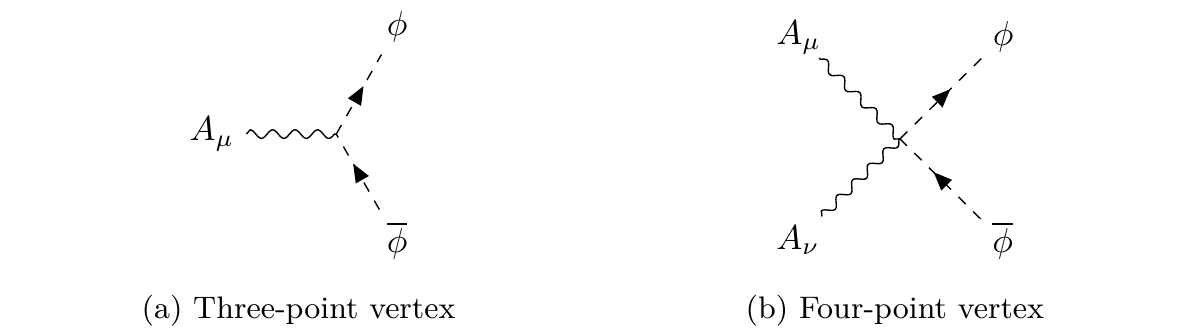}
\caption{\label{fig:scalar_vertices}Two vertices given by scalar HIP-photon interactions.}
\end{figure}

The production of spin-0 HIPs in the magnetic-dual theory of scalar
QED is described by the candidate Lagrangian
\begin{equation}
\label{eq:scalarLagrangian}
\mathcal{L} = -\frac{1}{4}F^{\mu \nu} F_{\mu \nu} + (D^\mu \phi)^\dagger 
(D_\mu \phi) - m^2 \phi^\dagger \phi,
\end{equation}
where the covariant derivative
\begin{equation}
\label{eq:covDerivative}
D_\mu = \partial_\mu + igA_\mu
\end{equation}
couples the spin-0 HIP field, $\phi$, to the photon field, $A_\mu$, whose field
strength is $F_{\mu \nu} = \partial_\mu A_\nu - \partial_\nu A_\mu$.
As a result, the covariant derivative component of 
\ref{eq:scalarLagrangian} gives rise to the HIP interactions
\begin{equation}
 \mathcal{L}_\text{int} = \mathcal{L}_\text{3pt} + \mathcal{L}_\text{4pt},
\end{equation}
whose components
\[\mathcal{L}_\text{3pt} = -igA_\mu (\phi^\dagger \partial^\mu \phi - \phi 
\partial^\mu \phi^\dagger)\]
and
\[\mathcal{L}_\text{4pt} = g^2 A^\mu \phi^\dagger A_\mu \phi\]
dictate the vertices shown in figure~\ref{fig:scalar_vertices}.

\begin{figure}[!h]
    \centering
    \includegraphics{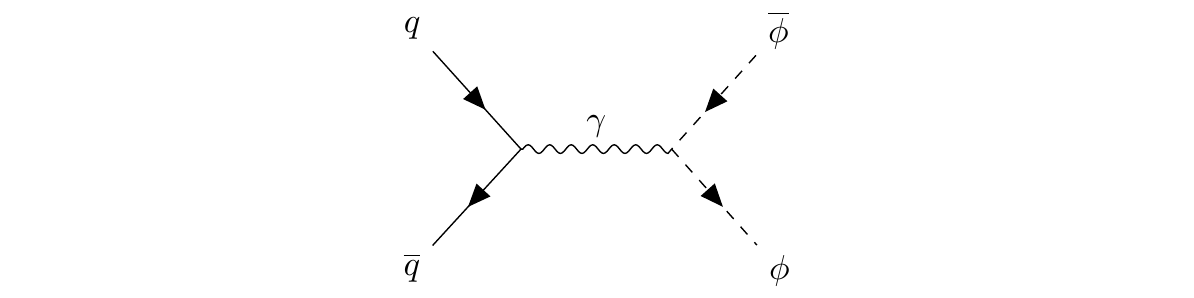}
    \caption{\label{fig:scalar_DYdiagram}Feynman diagram of the photon-exchange Drell-Yan mechanism to produce a pair of scalar HIPs.}
\end{figure}

\begin{figure}[!h]
\centering
\includegraphics{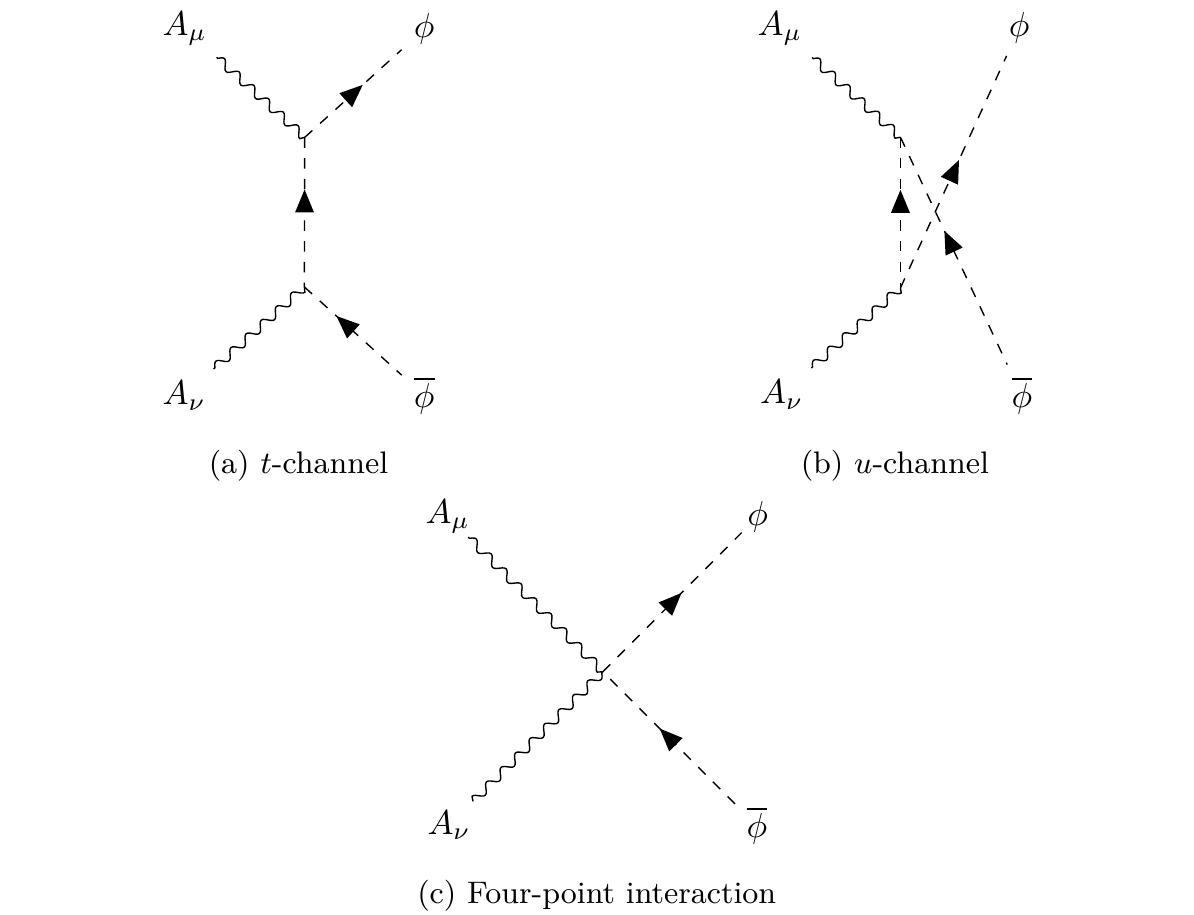}
\caption{\label{fig:scalar_PFdiagrams}Feynman diagrams of photon fusion to produce a pair of scalar HIPs.}
\end{figure}

While the Drell-Yan mechanism for scalar HIP pair production, shown in
figure~\ref{fig:scalar_DYdiagram}, realizes only the three-point vertex,
the photon-fusion process, shown in figure~\ref{fig:scalar_PFdiagrams}, has
contributions from both the three-point and four-point vertices.

\begin{figure}[!h]
\centering
\includegraphics{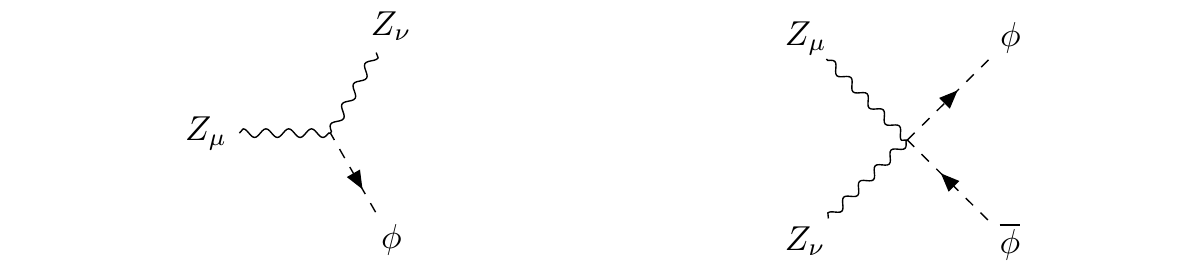}
\caption{\label{fig:scalarHECO_Z}Feynman diagrams of tri-linear
      (left) and quadra-linear (right) interactions of spin-0 HECOs
      with $Z^0$ bosons.}
\end{figure}

The interaction between the spin-0 HECOs and the $Z^0$ boson can be
derived through the gauge invariance principle. Spin-0 HECOs couple to
the $Z^0$ boson
via tri-linear and quadra-linear interactions, therefore, they can only 
interact with a pair of $Z^0$ bosons assuming Standard-Model interactions, as 
shown in figure~\ref{fig:scalarHECO_Z}. That is, the spin-0 HECOs cannot be 
produced via the $Z^0$-exchange-mediated Drell-Yan mechanism.

\subsection{Spin-\textonehalf\ interaction} 
\begin{figure}[!h]
    \centering
    \includegraphics{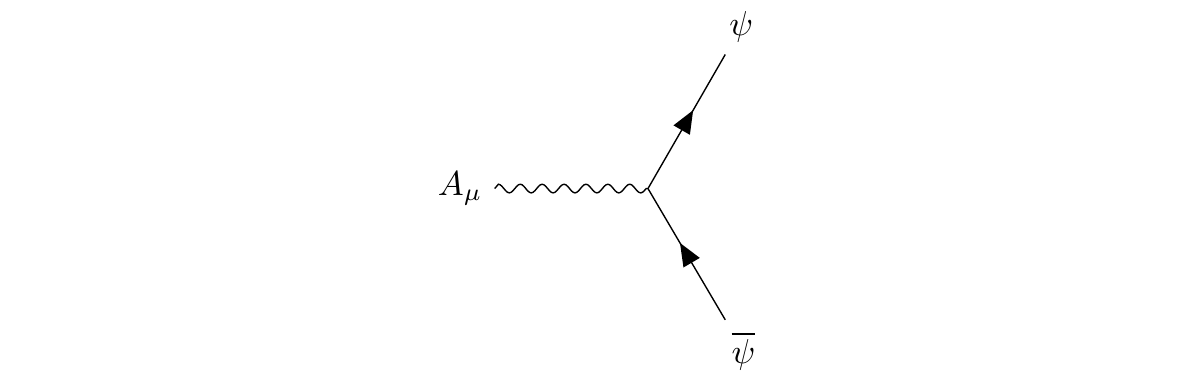}
    \caption{\label{fig:fermion_photonVertex}Three-point vertex 
     for the interaction of spin-\textonehalf\ HIPs with a photon.}
\end{figure}

\begin{figure}[!h]
    \centering
    \includegraphics{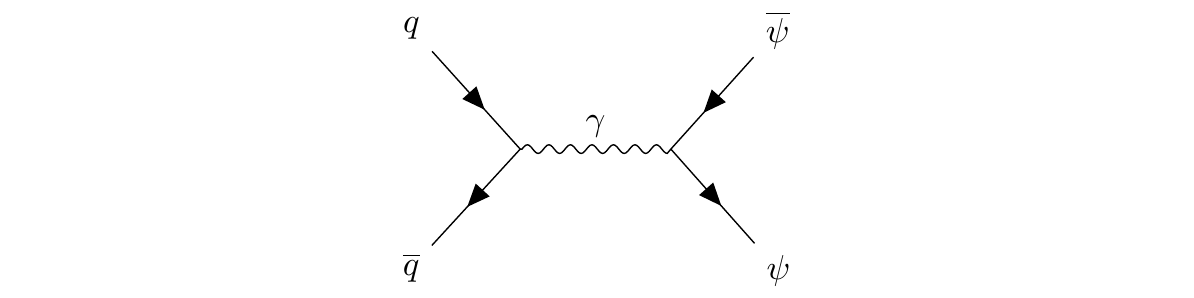}
    \caption{\label{fig:fermion_DYdiagram}Feynman diagram of the
      photon-exchange Drell-Yan mechanism to produce a pair of
      fermionic HIPs.}
\end{figure}

\begin{figure}[!h]
\centering
\includegraphics{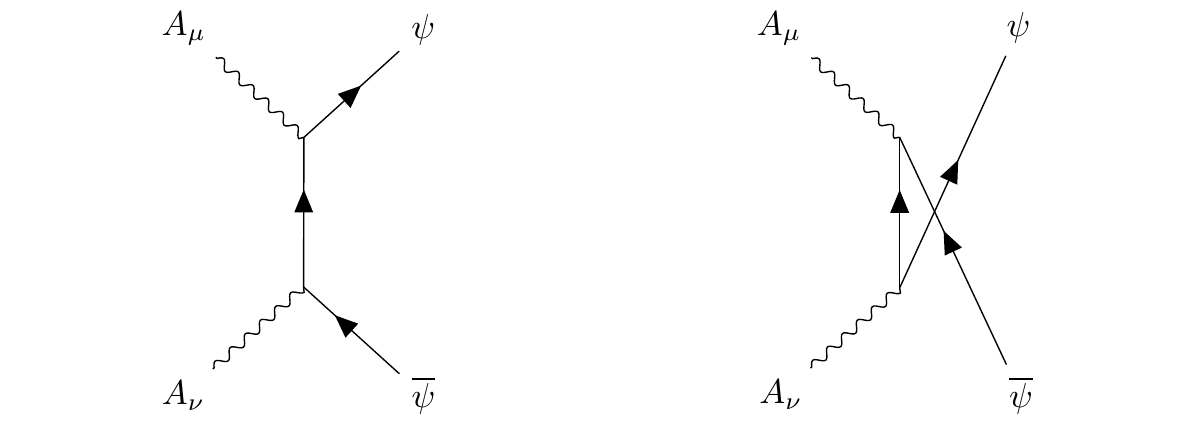}
\caption{\label{fig:fermion_PFdiagrams}Feynman diagrams of photon
      fusion to produce a pair of fermionic HIPs.}
\end{figure}

Assuming HIPs are Dirac fermions, the basic Lagrangian for a 
spin-\textonehalf\ field $\psi$ is
\[\mathcal{L} = -\frac{1}{4}F^{\mu \nu} F_{\mu \nu} + 
\overline{\psi} i \gamma^\mu D_\mu \psi - m\overline{\psi} \psi,\]
with a different covariant derivative component of 
$\overline{\psi} i \gamma^\mu D_\mu \psi$ compared to that of the spin-0 
scenario. This distinction gives
rise to a different HIP-photon interaction term
\begin{equation}
\label{eq:EMInteraction}
 \mathcal{L}_\text{int} = - g\overline{\psi} \gamma^\mu A_\mu \psi,
\end{equation}
which corresponds to the three-point vertex of spin-\textonehalf\ HIPs in 
figure~\ref{fig:fermion_photonVertex}. 
With this vertex, the photon-mediated Drell-Yan mechanism in
figure~\ref{fig:fermion_DYdiagram} was the first production process
considered for HIP production at the LHC. In addition, this three-point vertex
contributes to the photon-fusion production process in 
figure~\ref{fig:fermion_PFdiagrams}, where two such vertices are involved for 
a given production channel.

\begin{figure}[!h]
    \centering
    \includegraphics{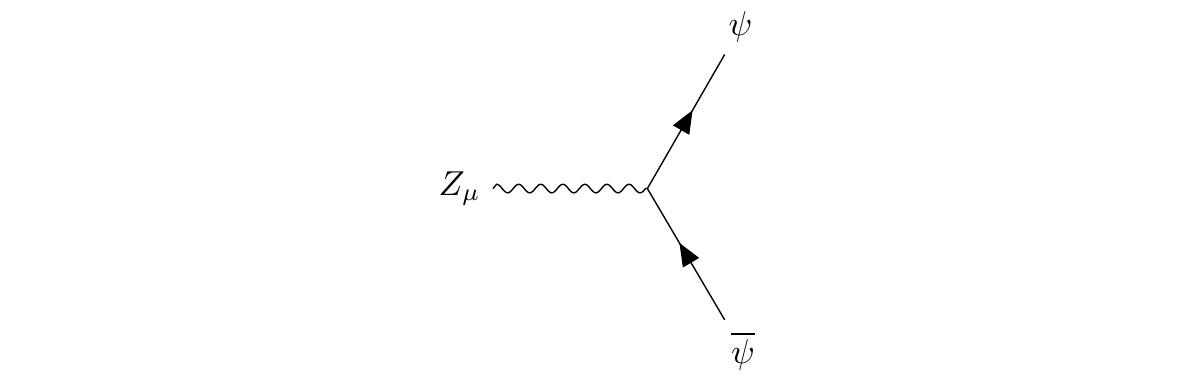}
    \caption{\label{fig:fermion_ZVertex}Three-point vertex for the
      interaction of spin-\textonehalf\ HECOs with a $Z^0$ boson.}
\end{figure}

\begin{figure}[!h]
    \centering
    \includegraphics{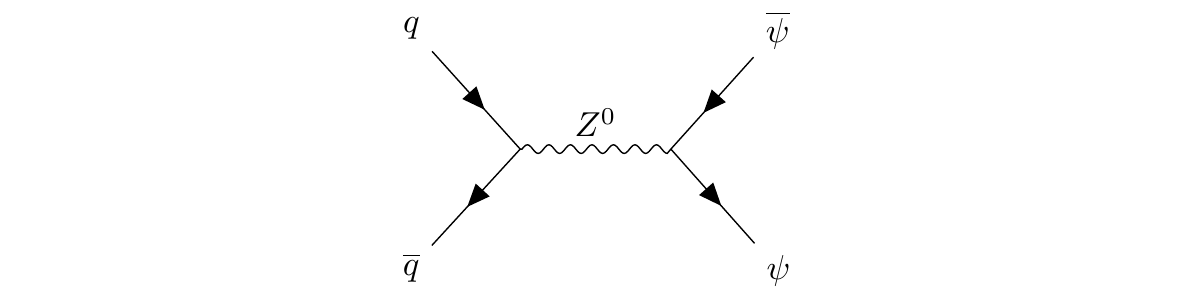}
    \caption{\label{fig:fermion_DYdiagram_Z}Feynman diagram of the $Z^0$-exchange Drell-Yan mechanism to produce a pair of fermionic HECOs.}
\end{figure}

While the magnetically charged monopoles are only allowed to interact
via the electromagnetic neutral current, the electrically charged
HECOs, like many other particles with exotic electric
charges~\cite{Kang_2008,Langacker_2011}, can also interact via the
weak neutral current, as we assume they transform under the $SU(2)$
symmetry of the weak interactions.  To build such a model, a local
$SU(2)$ gauge invariance is imposed to generate a coupling to $Z^0$
bosons via a covariant derivative analogous to~\ref{eq:covDerivative}. 
This interaction is given by
\begin{equation}
\label{eq:weakInteraction}
    \mathcal{L}'_\text{int} = -\frac{e}{2\sin \theta_\text{W} \cos \theta_\text{W}} \overline{\psi} \gamma^\mu (c_\text{L} P_\text{L} + c_\text{R} P_\text{R})\psi Z_\mu,
\end{equation}
where $e$ is the electromagnetic coupling, $\theta_\text{W}$ is the weak
mixing angle, and $P_\text{L}$ and $P_\text{R}$ are the chiral projectors. 
Unlike the HIP-photon interaction in~\ref{eq:EMInteraction}, this Lagrangian
has a chirality dependence parametrized by two couplings 
that are generally not identical:
\begin{equation}
\label{eq:chiralCouplings}
    c_\text{L} = t_3 - |z| \sin^2 \theta_\text{W} \quad \text{and} \quad
    c_\text{R} = - |z| \sin^2 \theta_\text{W},
\end{equation}
where $t_3$ is the weak isospin of the HIPs. The assumption that HECOs
transform as $SU(2)$ singlets implies that $t_3 = 0$ for HECOs. The
interaction Lagrangian in~\ref{eq:weakInteraction} is associated
with the three-point vertex in figure~\ref{fig:fermion_ZVertex}, which
drives the Drell-Yan production of spin-\textonehalf\ HECOs via
$Z^0$-boson exchange, as in figure~\ref{fig:fermion_DYdiagram_Z}.

\section{\label{sec:heavyIon}Ultraperipheral heavy-ion collisions}

Heavy-ion collisions have many distinctive properties that make them a
unique environment for studying a wide range of interactions,
including the production of magnetic monopoles.  The impact parameter
of the colliding ions characterizes the effective interaction in
heavy-ion collisions. When the ions interact with impact parameters
less than or equal to twice the nuclear radii, the strong force
governs the interaction between them. These are known as central
collisions and their extraordinarily high temperatures can produce a
quark-gluon plasma. However, when the two ions pass each other
with an impact parameter much larger than twice their radii, strong
interactions between the nucleons are no longer
possible. Nevertheless, even at large impact parameters
electromagnetic processes between the ions can occur. In these
so-called ultraperipheral collisions (UPC), photons radiated from the
ions interact via the photon-fusion process while the ions remain
intact.

The prospect of monopole production in heavy-ion collisons has a long
history of interest. Initially, it was envisioned to occur via thermal
processes of central collisions where the massive monopoles would
couple to the quark-gluon plasma~\cite{Roberts1986DiracMP} via
photons. Later, monopole production was postulated in ultraperipheral
elastic photon-fusion processes in electron-electron, proton-proton
and lead-lead collisions~\cite{Reis_2017}, motivated by the enhanced cross
sections due to the high electric charges of the heavy ions over protons. 
Recently, monopole production has been considered again in ultraperipheral 
heavy-ion collisions~\cite{Gould:2017fve,Gould:2017zwi,Gould_2019,Rajantie_2019,Ho:2019ads,Ho:2021uem,Gould:2021bre}, where the strong magnetic
field from the ultrarelativistic high-charge ions passing each other may produce monopoles via the magnetic-dual mechanism of the
thermal Schwinger
process~\cite{Schwinger_1951}. This production mechanism has been considered in
an experimental search for the first time by the MoEDAL 
collaboration~\cite{MoEDAL2021Schwinger}. 
The present study first focuses on the
electromagnetic radiation emitted by the ultrarelativistic ions, as in
\cite{Reis_2017}, then continues to formulate a model of monopole and HECO
production in (the magnetic dual of) QED, assuming the effective interaction 
is photon fusion.

\subsection{Electromagnetic radiation emitted by relativistic heavy ions}

We review the \textit{equivalent photon method}, which describes the electromagnetic radiation emitted in heavy-ion collisions. The original formalism, proposed by Enrico Fermi in 1924~\cite{Fermi1924}, treated the time-varying electromagnetic field as a flux of virtual photons. Later, it was adapted by Weizs\"acker and Williams~\cite{VonWeizsacker,Williams} for the photon fluxes of relativistic ions.

We consider a particle with electric charge $Z$ as approximately a
point particle. At ultrarelativistic velocities, it produces electric and 
magnetic fields ($\mathbf{E}$ and $\mathbf{B}$) such that the total energy 
they carry propagates as a flux of equivalent photons with its direction 
and magnitude dictated by $\mathbf{E} \times \mathbf{B}$. The resulting 
photon density spectrum $dn_\gamma$ possesses a $Z^2$ dependence and can 
be defined for a given pair of energy $\omega$ and impact parameter $b$   
in the plane transverse to its direction of motion.
The number density over the entire transverse plane is obtained by 
integrating over all possible impact parameters $b$. Therefore, 
the number density per energy is
\begin{equation}
\label{eq:defPhotonFlux}
    \frac{dn_\gamma}{d\omega} (\omega) = 2\pi \int_{b_\text{min}}^\infty \frac{dn_\gamma}{d\omega} (\omega, b) b db.
\end{equation}
Defined with respect to one particle, this impact parameter, denoted as $b$
hereafter, should be distinguished from the collision impact parameter that
categorizes the collision type. If the two colliding particles have the same 
size, the particle impact parameter $b$ is half of the collision impact
parameter on average.

Now we consider the scenario of two charged particles moving towards
each other at ultrarelativistic velocities.  In an ultraperipheral
collision, the nuclei remain intact and the interactions take place at
a finite distance, which in turn restricts the kinematics of the
virtual photons due to the uncertainty principle.  In consequence, the
emitted photon's wavelength given by the impact parameter $b$ has a
lower bound of the nuclear radius $R_A$, which corresponds to a
maximal transverse momentum of $\hbar/R_A$.  Since the photon's
off-shell mass is suppressed by the impact parameter, its virtuality
decreases with the size of the charged particle. In the longitudinal
direction, the photon's momentum is boosted by a Lorentz factor,
reaching an effective maximum of $\gamma \hbar/R_A$. It follows from
the $1/R_A$ dependence that the photon energy bound is more stringent
in heavy-ion collisions than in proton-proton collisions.

\section{\label{sec:collisions}Proton-proton and heavy-ion collisions at the LHC}

Theories describing the production of magnetic monopoles and HECOs
provide little guidance as to their masses. However, when it comes to
collider searches, the accessible mass range is bounded by the
collision energy. At the LHC collision energy, it is unlikely to 
produce HIPs more massive than 4000~GeV via either the Drell-Yan or 
the photon-fusion mechanism, given that
the interacting partons only carry a fraction $x$ of the beam energy.
In ultraperipheral heavy-ion collisions, the even smaller
accessible mass range is
restricted by the maximum energy of the emitted photon, $\gamma \hbar/R_A$, which is 82~GeV for lead-lead collisions.

Table~\ref{tab:parametersLHC} compares the parameters of
ultraperipheral collisions between different particles at the LHC design beam 
energy, where the ions are fully stripped with no remaining electrons. 
The proton-proton collision center-of-mass energy $\sqrt{s}$ is equal
to the sum of the proton beam energies, i.e., 13~TeV in Run~2 at the
LHC. The heavy-ion center-of-mass energy $\sqrt{s_\text{NN}}$ is
determined by the energy carried by the protons in the colliding
beams, since only the protons in the ions are accelerated. As a
consequence, given the same beam energy, $\sqrt{s_\text{NN}}$ for
heavy-ion collisions is lower than $\sqrt{s}$ for proton-proton
collisions. 
Given the $1/R_A$ dependence of the 
maximal transverse momenta $p_\text{T}^\text{max}$
for the different collision systems, the $p_\text{T}$ profile of the photons
is highly
suppressed in a heavy-ion collision compared 
to a proton-proton collision.

\begin{table}[!h]
\renewcommand{\arraystretch}{1.25} 
\begin{center}
  \caption{\label{tab:parametersLHC}Parameters in different ultraperipheral collision systems with the same beam energy~$E_\text{beam}$: nucleon-nucleon center-of-mass energy~$\sqrt{s_\text{NN}}$, Lorentz factor~$\gamma$, nuclear radius~$R_A$, charge-induced photon luminosity enhancement~$Z^4$, charge-to-mass ratio~$Z/A$, maximal transverse momentum $p_\text{T}^\text{max}$, maximal longitudinal momentum $p_\text{L}^\text{max}$ and maximal photon-photon center-of-mass energy $\sqrt{s_{\gamma \gamma}^\text{max}}$.}
  \begin{tabularx}{0.95\textwidth}{ c *{9}{Y} }
    \toprule
    \toprule
    {} & {$E_\text{beam}$ [TeV]} & {$\sqrt{s_\text{NN}}$ [TeV]} & {$\gamma$} & {$R_A$ [fm]} & {$Z^4$} & {$Z/A$} & {$p_\text{T}^\text{max}$ [MeV]} & {$p_\text{L}^\text{max}$ [GeV]} & {$\sqrt{s_{\gamma \gamma}^\text{max}}$ [GeV]}\\
    \midrule
     p & 7.0 & 14 & $7.5 \times 10^3$ & 0.8~\cite{ProtonRadius} & 1.0 & 1.0 & 237 & $1.8 \times 10^3$ & $3.6 \times 10^3$\\
     \ce{^{16}_{8}O} & 7.0 & 7.0 & $3.8 \times 10^3$ & 3.0 & $4.1 \times 10^3$ & 0.50 & 65 & 245 & 490 \\
     \ce{^{129}_{54}Xe} & 7.0 & 5.9 & $3.1 \times 10^3$ & 6.1 & $8.5 \times 10^6$ & 0.42 & 32 & 102 & 204 \\
     \ce{^{208}_{82}Pb} & 7.0 & 5.5 & $3.0 \times 10^3$ & 7.1 & $4.5 \times 10^7$ & 0.39 & 28 & 82 & 164 \\
    \bottomrule
    \bottomrule
  \end{tabularx}
\end{center}
\end{table}

The LHC has a long-standing interest in collisions between lighter ions, 
which have an advantage in ultraperipheral collisions.
In Run~2, the LHC delivered xenon-xenon collisions in addition to 
lead-lead collisions.
Different collision systems have different ratios of
atomic number $Z$ and mass number $A$. Hence, $\sqrt{s_\text{NN}}$ is
higher for an ion with a higher ratio of atomic number $Z$ to mass
number $A$, for a fixed beam energy.
Heavy nuclei with 
high $Z$ values have significantly stronger electrostatic interactions 
between the protons extending beyond their sizes. Hence, higher fractions 
of neutrons are necessary for the short-range strong interaction between 
nucleons to keep the ion stable. In other words, light ions such as 
oxygen have higher $Z/A$ values than the commonly considered xenon 
and lead ions. The energy fraction carried by the protons will increase 
and result in higher photon-photon collisions energies. The LHC foresees the 
addition of oxygen-oxygen collisions to its heavy-ion collision program in 
Run 3.

\section{\label{sec:computation}MadGraph implementation and results}

The HIP production model is built within an event generator called
\textsc{MadGraph}5\_aMC@NLO~\cite{MadGraph}. The model itself defines
particles and their interactions that are passed to the
\textsc{MadGraph}5\_aMC@NLO generator for every user-specified process
of interest. Next, \textsc{MadGraph}5\_aMC@NLO searches for all 
interaction vertices in the specified model that accommodate 
the incoming and outgoing particles the user provides to 
generate a valid hard process through a set of Feynman diagrams.  
It then calculates the matrix element of the hard process
from the Feynman diagrams via perturbation theory to a finite
order. It outputs the production cross section as an integral of the
matrix element over the initial momenta of the incoming particles and
the phase space of the outgoing particles.  In hadron collisions, the
probability of finding an incoming particle or a parton at a given
momentum follows the parton distribution function (PDF). The
\textsc{MadGraph}5\_aMC@NLO generator provides different PDF options 
via the \verb|LHAPDF| library. 

\subsection{Couplings}
Every HIP interaction vertex described in section~\ref{sec:production} is
characterized by a coupling constant that is proportional to the HIP charge.
\textsc{MadGraph}5\_aMC@NLO
uses Heaviside-Lorentz units such that one electron charge is $\sqrt{4\pi}$
times the electron charge in the Gaussian units. Thus the fine structure
constant is written as $\alpha = \frac{e^2}{4\pi}$, 
where the electron coupling $e$ in Heaviside-Lorentz units is 
\begin{equation}
\label{eq:ge}
e = 2 \sqrt{\alpha \pi}.
\end{equation} 
As a result, the HECO couplings are integer multiples of the electron
coupling, while the monopole couplings are integer multiples of $g_\textrm{D}$,
obtained by rearranging~\ref{eq:ChQuantization} and substituting~\ref{eq:ge}:
\begin{equation}
\label{eq:gD}
g_\textrm{D} = \frac{2\pi}{\sqrt{4\pi \alpha}} = \sqrt{\frac{\pi}{\alpha}}.
\end{equation}

\subsection{Spin-0 and spin-\textonehalf\ photon-HIP vertices}
The spin-0 HIP-photon interaction features the two vertices in
figure~\ref{fig:scalar_vertices}, where the three-point vertex is needed
for both the Drell-Yan and the photon-fusion mechanisms while the
four-point vertex is only relevant for photon fusion. On the contrary,
the spin-\textonehalf\ HIP's interactions with a photon only feature
the three-point vertex, which is responsible for both the Drell-Yan
induced and the photon-fusion induced HIPs. For a given spin
scenario, the two different partonic mechanisms can be validated
through the validation of the same vertices. For
each vertex, it suffices to validate one interaction that involves
it. Since the photon-fusion mechanism utilizes all three vertices,
including the additional spin-0 four-point vertex, a successfully
constructed photon-fusion mechanism guarantees the validity of its
Drell-Yan counterpart. Hence, we validate the photon-HIP
vertices in photon-fusion interactions.

Since the cross section depends explicitly on the center-of-mass
energy, it is ideal to work with collisions with a fixed
center-of-mass energy. However, the momenta of the photons emitted by
two colliding hadrons follow a non-trivial distribution such that the
effective photon collisions have no fixed center-of-mass energy. For
the ease of the theoretical calculations, the photon-fusion process is
modeled in \textsc{MadGraph}5\_aMC@NLO by colliding \textit{bare
  photons}, i.e., photons that are not emitted by hadrons.
In this case, the initial momenta of the photons are automatically set
to a constant value given the fixed collision energy. Cross sections
are obtained for the photon collisions generated with
\textsc{MadGraph}5\_aMC@NLO and calculated analytically to validate
the vertex implementation. Given all the vertices discussed in
section~\ref{sec:production}, the production cross sections at tree level 
are computed to be
\begin{equation}
\label{eq:xsection_spin0}
\sigma (\gamma \gamma \to \phi \overline{\phi}) = \frac{g^4}{64\pi}
\left [ \frac{2\beta}{E^2}  + \frac{2\beta m^4}{E^6} +  \frac{E^2 (1 -
    \beta^2)^2 - 2m^2}{E^4} \ln \left( \frac{1 + \beta}{1-\beta}
  \right) \right ],  \quad \text{for spin-0 HIPs, and}
\end{equation}
\begin{equation}
\label{eq:xsection_spin1/2}
\sigma (\gamma \gamma \to \psi \overline{\psi}) = \frac{g^4}{32\pi E^2} \left [ 2\beta^3 - 4\beta + \frac{2E^4 + 2E^2 m^2 - m^4 }{E^4} \ln \left( \frac{1+\beta}{1-\beta}\right) \right], \quad \text{for spin-\textonehalf\ HIPs},
\end{equation}
where $E$ is the energy of each photon, $g$ stands for the coupling strength, $\beta$ is the normalized velocity and $m$ is the rest mass for each HIP.

The photon-fusion cross sections are generated by the
\textsc{MadGraph}5\_aMC@NLO model with two colliding photons, 
each carrying a fixed energy of 7~TeV. Clearly, the 14~TeV center-of-mass 
energy for the two-photon system is beyond the reach of the LHC. 
However, for validation purposes this energy can be arbitrarily chosen, 
as long as it is greater than the total mass of the final-state particles. 
The \textsc{MadGraph}5\_aMC@NLO model calculation is validated against 
theory computed from~\ref{eq:xsection_spin0} and~\ref{eq:xsection_spin1/2} 
for 1$g_\textrm{D}$ monopoles 
in table~\ref{tab:mono} and for $|z|=60$ HECOs (for their approximate 
equivalence to 1$g_\textrm{D}$ monopoles dictated by~\ref{eq:ChQuantization}) 
in table~\ref{tab:60e}.

\begin{table}[!h]
\begin{center}
  \caption{\label{tab:mono}Comparison of cross sections obtained from
    the model and calculated analytically for $|g|=1g_\textrm{D}$ monopoles of
    various mass-spin combinations produced via photon fusion (PF) at a
    fixed center-of-mass energy of 14~TeV.}
  \begin{tabularx}{\textwidth}{ c *{6}{Y} }
    \toprule
    \toprule
    \multirow{2}{*}{\shortstack{Mass \\(GeV)}} & \multicolumn{3}{c}{Spin-0 1$g_\textrm{D}$ monopole PF $\sigma$ [pb]} & \multicolumn{3}{c}{ Spin-\textonehalf\ 1$g_\textrm{D}$ monopole PF $\sigma$ [pb]}  \\
      & {Model} & {Theory} & {Ratio (M/T)} & {Model} & {Theory} & {Ratio (M/T)}\\
      \midrule
   10 & $1.370 \times 10^4$ & $1.370 \times 10^4$ & 1.000 &  $3.696 \times 10^5$ &  $3.685 \times 10^5$ & 1.000\\ 
   50 & $1.369 \times 10^4$ & $1.369 \times 10^4$ & 1.000 & $2.815 \times 10^5$ & $2.814 \times 10^5$ & 1.000\\ 
   100 & $1.367 \times 10^4$ & $1.367 \times 10^4$ & 1.000 &  $2.434 \times 10^5$ & $2.434 \times 10^5$ & 1.000\\ 
   500 & $1.328 \times 10^4$ & $1.327 \times 10^4$ & 1.001 & $1.560 \times 10^5$ & $1.560 \times 10^5$ & 1.000\\
   1000 & $1.238 \times 10^4$ & $1.238 \times 10^4$ & 1.000 & $1.196 \times 10^5$ & $1.196 \times 10^5$ & 1.000\\
   3000 & 7835 & 7836 & 1.000 & $6.600 \times 10^4$ & $6.602 \times 10^4$ & 1.000 \\
   5000 & 5455 & 5450 & 1.001 & $3.663 \times 10^4$ & $3.660 \times 10^4$ & 1.001 \\
    \bottomrule
    \bottomrule
  \end{tabularx}
\end{center}
\end{table}
\begin{table}[!h]
\begin{center}
  \caption{\label{tab:60e}Comparison of cross sections obtained from
    the model and calculated analytically for $|z|=60$ HECOs of
    various mass-spin combinations produced via photon fusion (PF) at a
    fixed center-of-mass energy of 14~TeV.}
  \begin{tabularx}{\textwidth}{ c *{6}{Y} }
    \toprule
    \toprule
    \multirow{2}{*}{\shortstack{Mass \\(GeV)}} & \multicolumn{3}{c}{Spin-0 $|z|=60$ HECO PF $\sigma$ [pb]} & \multicolumn{3}{c}{ Spin-\textonehalf\ $|z|=60$ HECO PF $\sigma$ [pb]}  \\
      & {Model} & {Theory} & {Ratio (M/T)} & {Model} & {Theory} & {Ratio (M/T)}\\
      \midrule
   10 & 9215 & 9213 & 1.000 & $2.486 \times 10^5$ &  $2.486 \times 10^5$ & 1.000\\ 
   50 & 9211 & 9208 & 1.000 & $1.894 \times 10^5$ & $1.892 \times 10^5$ & 1.001 \\ 
   100 & 9196 & 9196 & 1.000 & $1.637 \times 10^5$ & $1.637 \times 10^5$ & 1.000\\ 
   500 & 8930 & 8924 & 1.001 & $1.049 \times 10^5$ & $1.049 \times 10^5$ & 1.000 \\
   1000 & 8328 & 8325 & 1.000 & $8.046 \times 10^4$ & $8.042 \times 10^4$ & 1.000\\
   3000 & 5270 & 5271 & 1.000 & $4.439 \times 10^4$ & $4.441 \times 10^4$ & 1.000\\
   5000 & 3669 & 3666 & 1.001 & $2.464 \times 10^4$ & $2.462 \times 10^4$ & 1.001\\
    \bottomrule
    \bottomrule
  \end{tabularx}
\end{center}
\end{table}

\subsection{Spin-\textonehalf\ Drell-Yan $Z^0$-exchange vertex for HECOs}
The implementation of the $Z^0$ boson decay to a pair of
spin-\textonehalf\ HECOs is similar to that of 
the decay of a $Z^0$ boson to an electron-positron pair, apart from the
necessary modifications to represent the zero weak isospin and
$|z|$~\textgreater~1 electric charges of the HECOs. This vertex is
further examined with cross-section calculations in
\textsc{MadGraph}5\_aMC@NLO itself to look for the expected charge
dependence. The cross section $\sigma$ goes as the absolute square of
the matrix element $\mathcal{M}$, given by the interaction Lagrangian
in \ref{eq:weakInteraction}, which scales as the couplings
$c_\text{L/R}$ in \ref{eq:chiralCouplings}:
\begin{equation}
    \sigma \propto |\mathcal{M}|^2 \propto |c_\text{L} + c_\text{R}|^2 \propto |z|^2.
\end{equation}
That is, the cross section of the $Z^0$-exchange mode grows as $|z|^2$,
just as in the photon exchange mode. Cross sections for both modes are
listed for mass 1000~GeV HECOs in table~\ref{tab:Z_exchange_xsections} for a
range of charges: $|z|$ = 2 and 7 as considered in multi-charged 
particle searches~\cite{MCPATLAS2018,MCPATLAS2015,MCPATLAS2013,CMS2016,CMS2013a}, 
along with five HECO charges centered at $|z|$ = 60. The two exchange modes
destructively interfere as a result of the relative sign between the
two interactions parametrized in \ref{eq:EMInteraction} and
\ref{eq:weakInteraction}, hence, the total cross section is
smaller than the photon-exchange cross section alone. 
\begin{table}[!h]
\begin{center}
  \caption{\label{tab:Z_exchange_xsections}Cross sections for 1000~GeV spin-\textonehalf\ HECOs with varying charges in different Drell-Yan production modes.}
  \begin{tabularx}{0.7\textwidth}{ c *{4}{Y} }
    \toprule
    \toprule
    \multirow{2}{*}{\shortstack{Charge \\ $|z|$} } &
    \multicolumn{3}{c}{Drell-Yan $\sigma$ for 1000~GeV Spin-\textonehalf\ HECOs [pb]} \\
      & {$\gamma$ exchange} & {$Z^0$ exchange} & {$\gamma/Z^0$ exchange} \\
    \midrule
     2 & $2.191 \times 10^{-4}$ & $8.903 \times 10^{-5}$ & $1.968 \times 10^{-4}$ \\
     7 & $2.684 \times 10^{-3}$ & $1.091 \times 10^{-3}$ & $2.411 \times 10^{-3}$ \\
     20 & 0.02191 & $8.903 \times 10^{-3}$ & 0.01968 \\
     40 & 0.08765 & 0.03561 & 0.07871 \\
     60 & 0.1972 & 0.08013 & 0.1771 \\
     80 & 0.3506 & 0.1425 & 0.3148 \\
     100 & 0.5478 & 0.2226 & 0.4919 \\
    \bottomrule
    \bottomrule
  \end{tabularx}
\end{center}
\end{table}

\subsection{Parton distribution functions}

The total cross section $\sigma_{AB}$ of an arbitrary A-B collision
can be expressed as the hard parton-level sub-process cross section
$\hat{\sigma}_{ab}$ reweighted with the appropriate parton
distribution functions (PDF) $f_{a/A}$ and $f_{b/B}$ for the initial
state momentum fractions $x_a$ and $x_b$:
\begin{equation}
\label{eq:hadronic_xsection_formula}
\sigma_{AB} = \int dx_a dx_b f_{a/A}(x_a) f_{b/B}(x_b) \hat{\sigma}_{ab}.
\end{equation}
In proton-proton collisions, A and B represent the incoming protons
and the initial momentum fractions $x_i$ ($i=a, b$) are evaluated with
respect to the protons. If the partonic mechanisms consider different
initial states, as for Drell-Yan and photon fusion, then the parton
distribution functions $f_{a/A}$ and $f_{b/B}$ will be different.  In
\textsc{MadGraph}5\_aMC@NLO, the NNPDF23LO\_qed PDF~\cite{2013NNPDF} was used for
Drell-Yan production while the LUXqed17 PDF~\cite{2016LUXqed,2018LUXqed} 
was used for photon fusion, due to the intrinsic uncertainties of the PDFs. 
The quark and gluon PDFs are relatively well constrained, with uncertainties
typically of order 5\%~\cite{2017NNPDF_quark_uncertainty}, while the
photon PDF is weakly constrained due to the other processes present in
the LHC data. With uncertainties up to
50\%~\cite{2013NNPDF_photon_uncertainty}, the photon PDFs ultimately
lead to large uncertainties in the predicted cross sections of the
photon fusion process.  This disadvantage is specifically targeted in
the calculation of the LUXqed17 PDF, where electron-proton scattering
data was collected to determine the photon PDF. The uncertainties in
this measurement are significantly reduced to 1--2\%~\cite{2016LUXqed}
for a wide momentum range, as the momentum transfers in the scattering
processes are carried by photons emitted from the protons.

\subsection{Equivalent photon fluxes for ultraperipheral heavy-ion collisions}

In heavy-ion ultraperipheral collisions, the total cross section for
HIP production factorizes into the product of the parton-level cross
section convoluted with the equivalent photon spectra from the two
colliding ion beams. The parton-level interaction is the same photon
fusion as discussed for proton-proton collisions, while the photon
spectrum is uniquely parametrized for a given nuclear charge $Z$ that
follows \ref{eq:defPhotonFlux} in the Weizs\"acker-Williams
approach~\cite{Jackson},
\begin{equation}
f_\gamma(x) = \frac{Z^2 \alpha}{\pi} \frac{1}{x} [2x_i K_0 (x_i) K_1 (x_i) - x_i^2 (K_1^2(x_i) - K_0^2(x_i)],
\end{equation}
where $x_i=x m_N b_\text{min}$ builds in the Lorentz boost effect through
the atomic mass $m_N$ and excludes hadronic interactions via the
minimal impact parameter requirement $b_\text{min} = R_A$, while $K_{0,1}$
are the modified Bessel functions of the second kind of zero and first
order that encode longitudinal and transverse polarizations of the
spectrum. In principle, there is no constraint on the intactness of
the nuclei for an ultraperipheral collision to take place, as long as
there is no hadronic component in the same collision event. However,
in a typical calculation, a cut of the nuclear radius is applied to
the impact parameter to ensure the absence of hadronic activity. The
nuclear charge dependence of the photon spectrum scales as $Z^2$,
enhancing the cross section in heavy-ion collisions by $Z^4$ compared
to photon fusion in proton-proton collisions. The photons are
transversely polarized for relativistic ions, as parametrized by the
dominance of $K_1$ over $K_0$.

The implemented $K_{0,1}$ factors are polynomial approximations of
the true $K_{0,1}$ functions, with a precision on the order of
$10^{-7}$~\cite{Bessel}. For a lead ion with 82 protons and 126
neutrons, the atomic mass $m_N$ is set to 208 times the proton
mass. The radius $R_A$ is set to 6.64~fm.

\section{\label{sec:results}Results and conclusions}

We compared the cross sections and kinematics for spin-0 and
spin-\textonehalf\ production of monopoles and HECOs 
via photon fusion in
ultraperipheral collisions to those for the Drell-Yan and photon-fusion
mechanisms in proton-proton collisions for the LHC design collision
energy of $\sqrt{s}=14$~TeV. Since the photons emitted by the lead ion
beams are kinematically bounded by 82~GeV at the lead-lead
collision energy of $\sqrt{s_\text{NN}}=5.5$~TeV, the HIP mass range
accessible in ultraperipheral collisions is much lower than that of
proton-proton collisions. Hence, we examined $1g_\textrm{D}$ monopoles and 
$|z|$ = 60 HECOs with masses below 82~GeV. The implemented photon
spectrum acts as the photon PDF for ultraperipheral HIP production.

\begin{figure}[!h]
    \centering
    \begin{minipage}{0.45\textwidth}{
    \includegraphics[width=\textwidth,keepaspectratio=true]{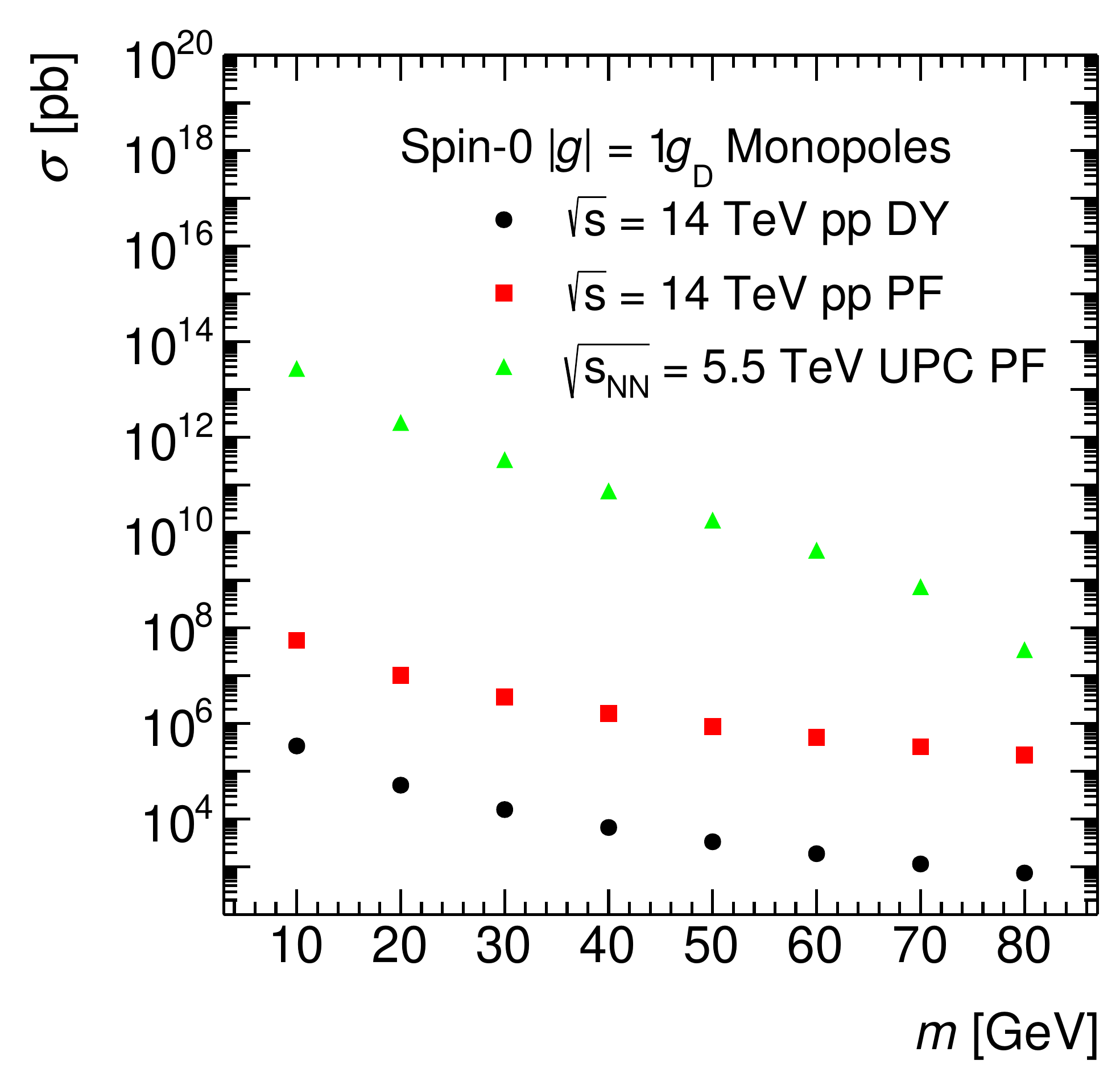}}\end{minipage}
\begin{minipage}{0.45\textwidth}{
    \includegraphics[width=\textwidth,keepaspectratio=true]{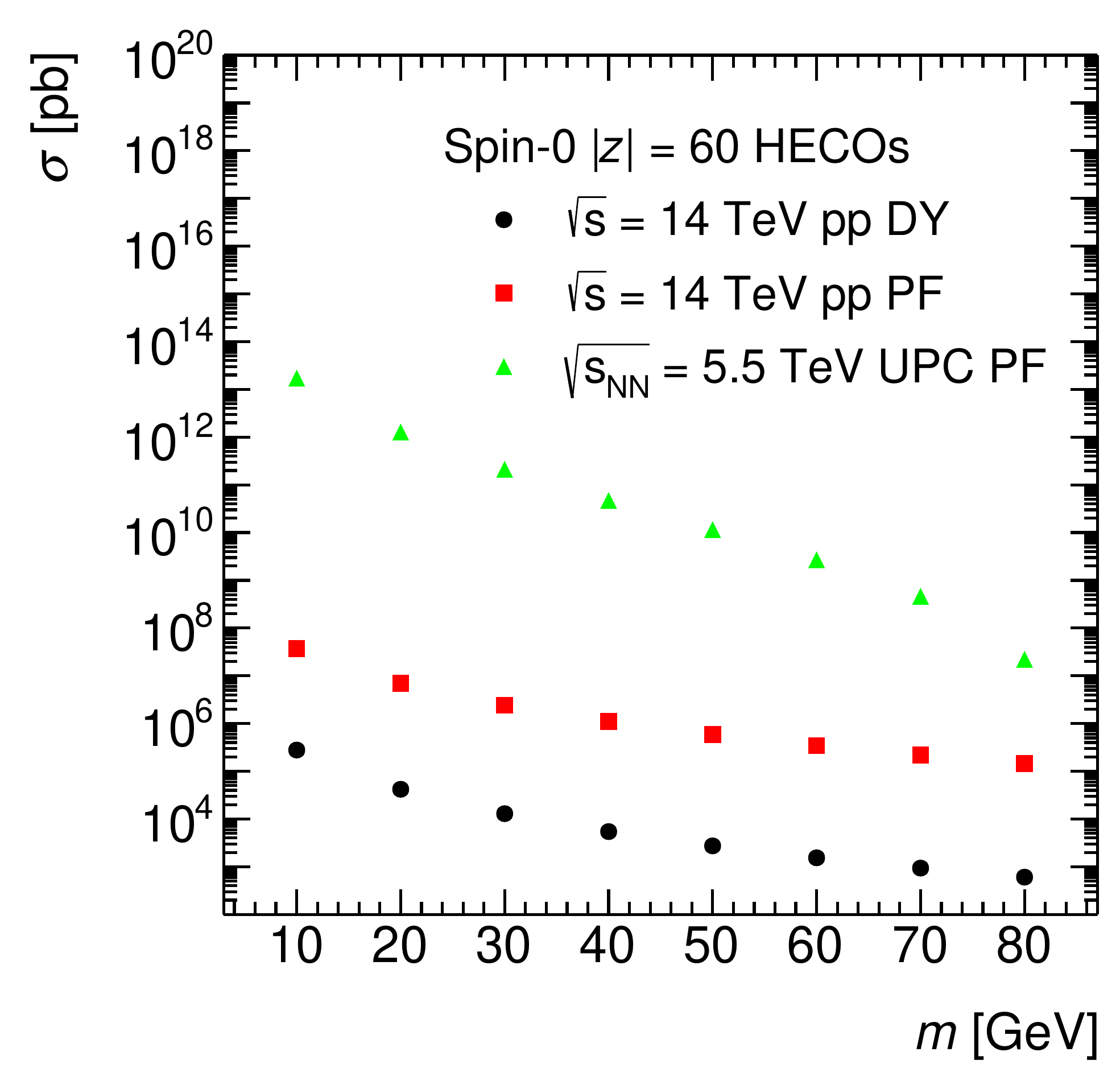}}\end{minipage}
    \caption{\label{fig:xsec_scalar}Cross sections for spin-0 HIPs via
      the Drell-Yan (DY) and photon-fusion (PF) mechanisms in proton-proton
      (pp) collisions and via photon fusion (PF) in lead-lead ultraperipheral
      collisions (UPC).}
\end{figure}

The production cross sections of spin-0 1$g_\textrm{D}$ monopoles and $|z| = 60$ HECOs 
are depicted in figure~\ref{fig:xsec_scalar}, where three different collision 
modes are presented. The lead-lead ultraperipheral production is 
compared to proton-proton production via the Drell-Yan and the photon-fusion 
mechanisms. The ultraperipheral production dominates the proton-proton 
production mechanisms at the considered mass range. As far as proton-proton 
collisions are concerned, the photon-fusion mechanism is dominant at
$\sqrt{s}=14$~TeV.

\begin{figure}[!h]
    \centering
    \begin{minipage}{0.45\textwidth}{
    \includegraphics[width=\textwidth,keepaspectratio=true]{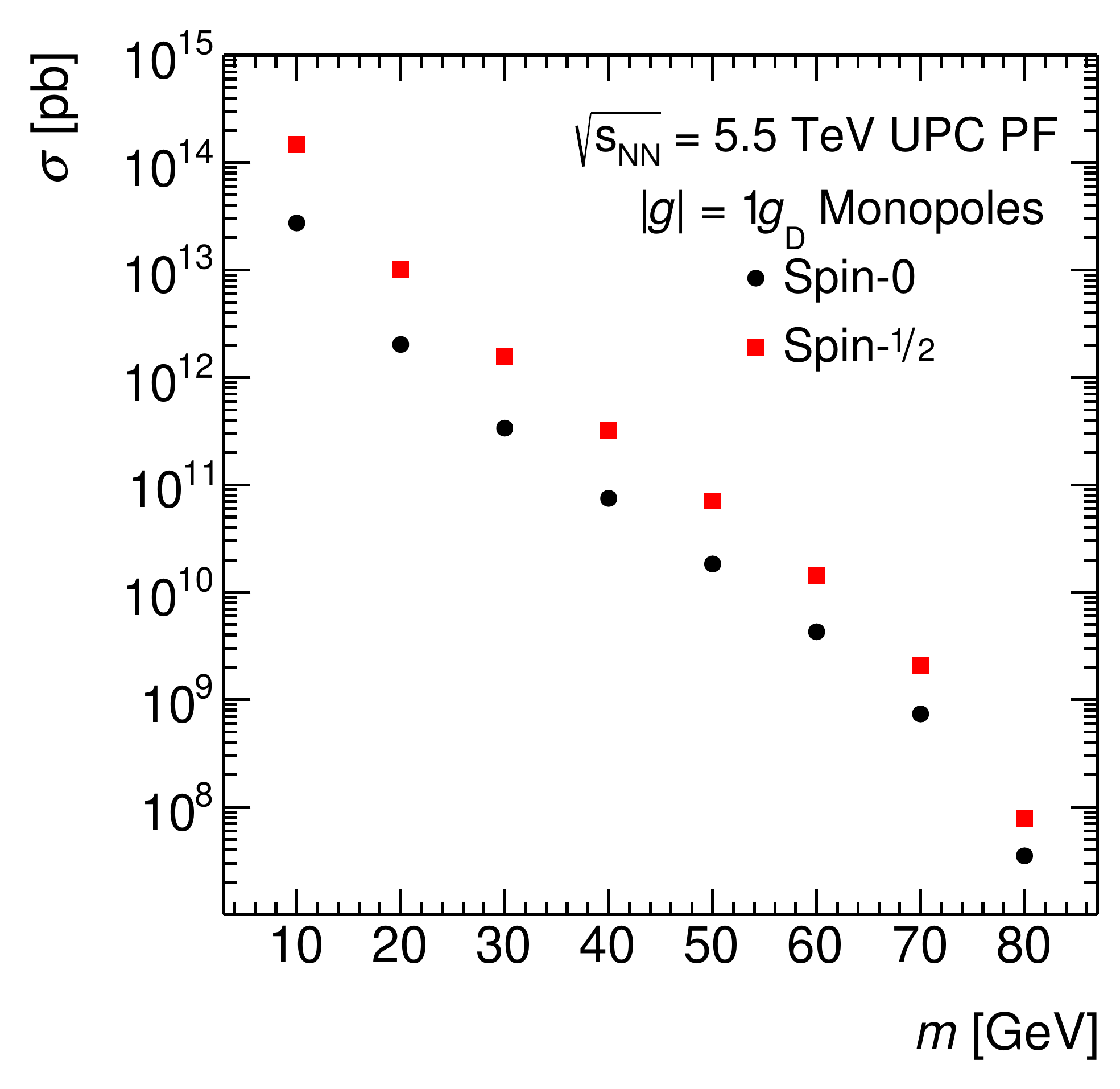}}\end{minipage}
\begin{minipage}{0.45\textwidth}{
    \includegraphics[width=\textwidth,keepaspectratio=true]{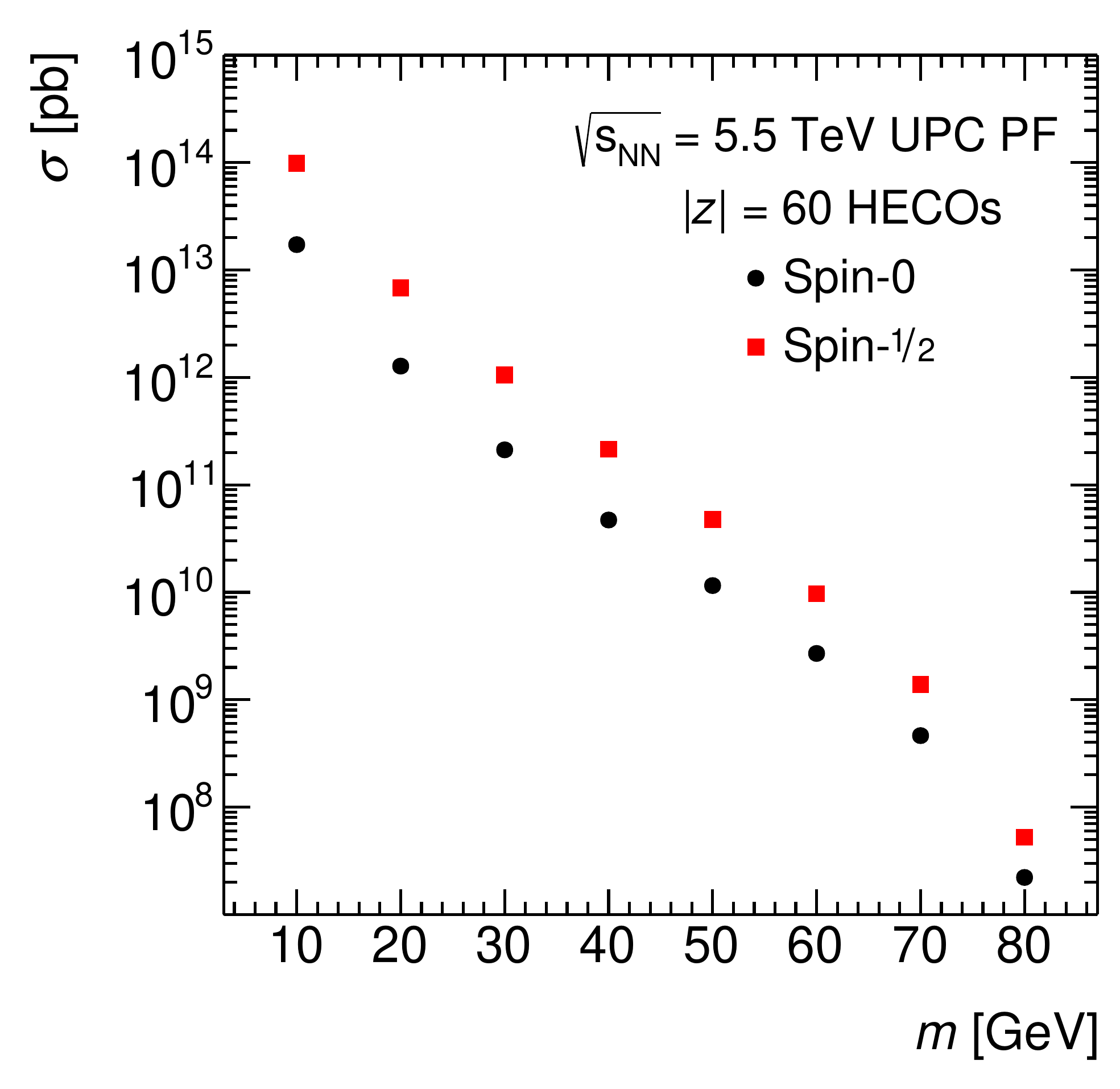}}\end{minipage}
    \caption{\label{fig:xsec_upc}Cross sections for spin-0 and
      spin-\textonehalf\ HIPs via photon fusion (PF) in lead-lead
      ultraperipheral collisions (UPC).}
\end{figure}

The ultraperipheral production cross sections of 1$g_\textrm{D}$ monopoles 
and $|z| = 60$ HECOs are shown in figure~\ref{fig:xsec_upc}, where the 
spin-\textonehalf\ HIPs have larger cross sections than the spin-0 HIPs.

\begin{figure}[!h]
\centering
\begin{minipage}{0.45\textwidth}{
    \includegraphics[width=\textwidth,keepaspectratio=true]{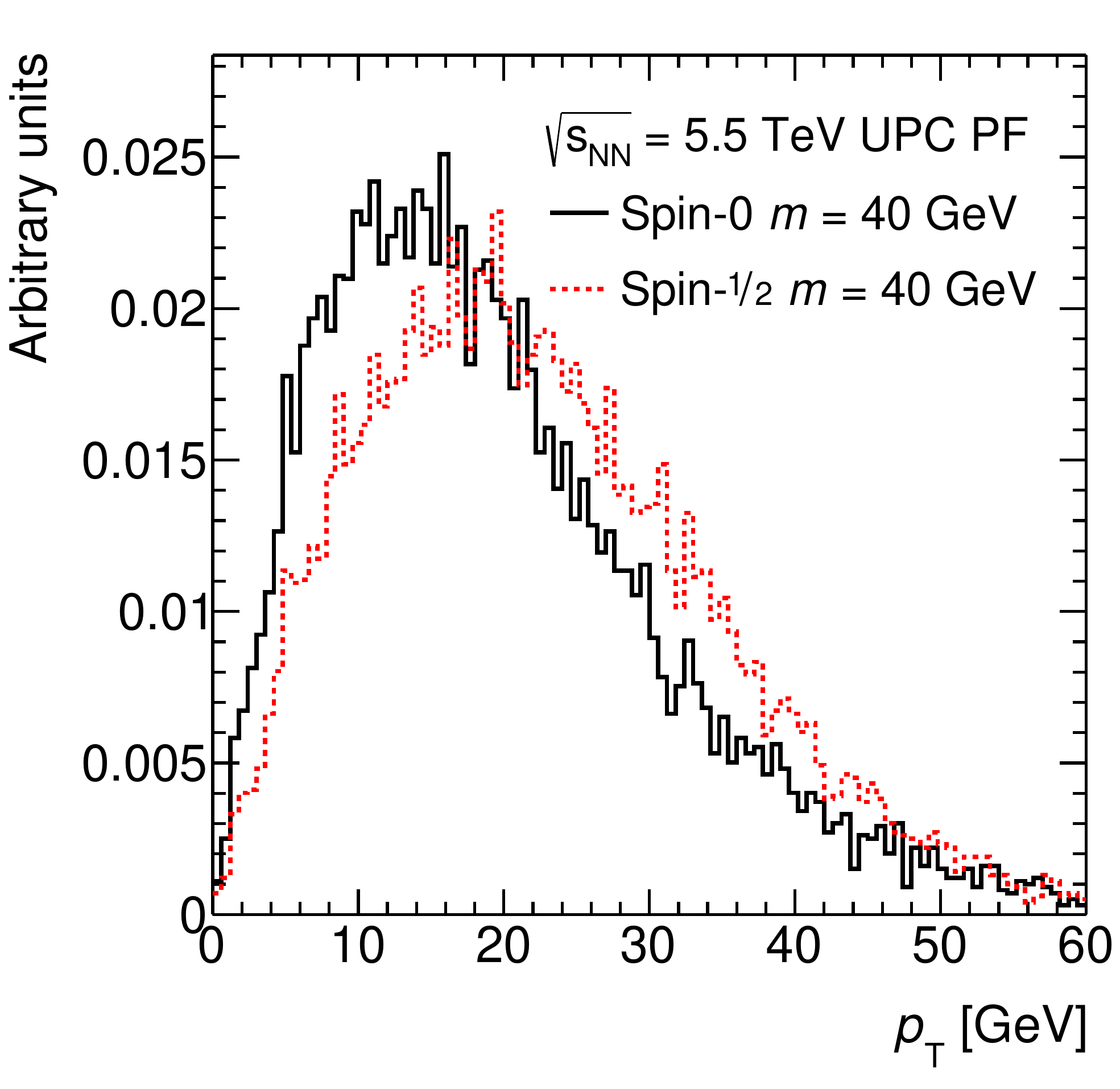}}\end{minipage}
\begin{minipage}{0.45\textwidth}{
    \includegraphics[width=\textwidth,keepaspectratio=true]{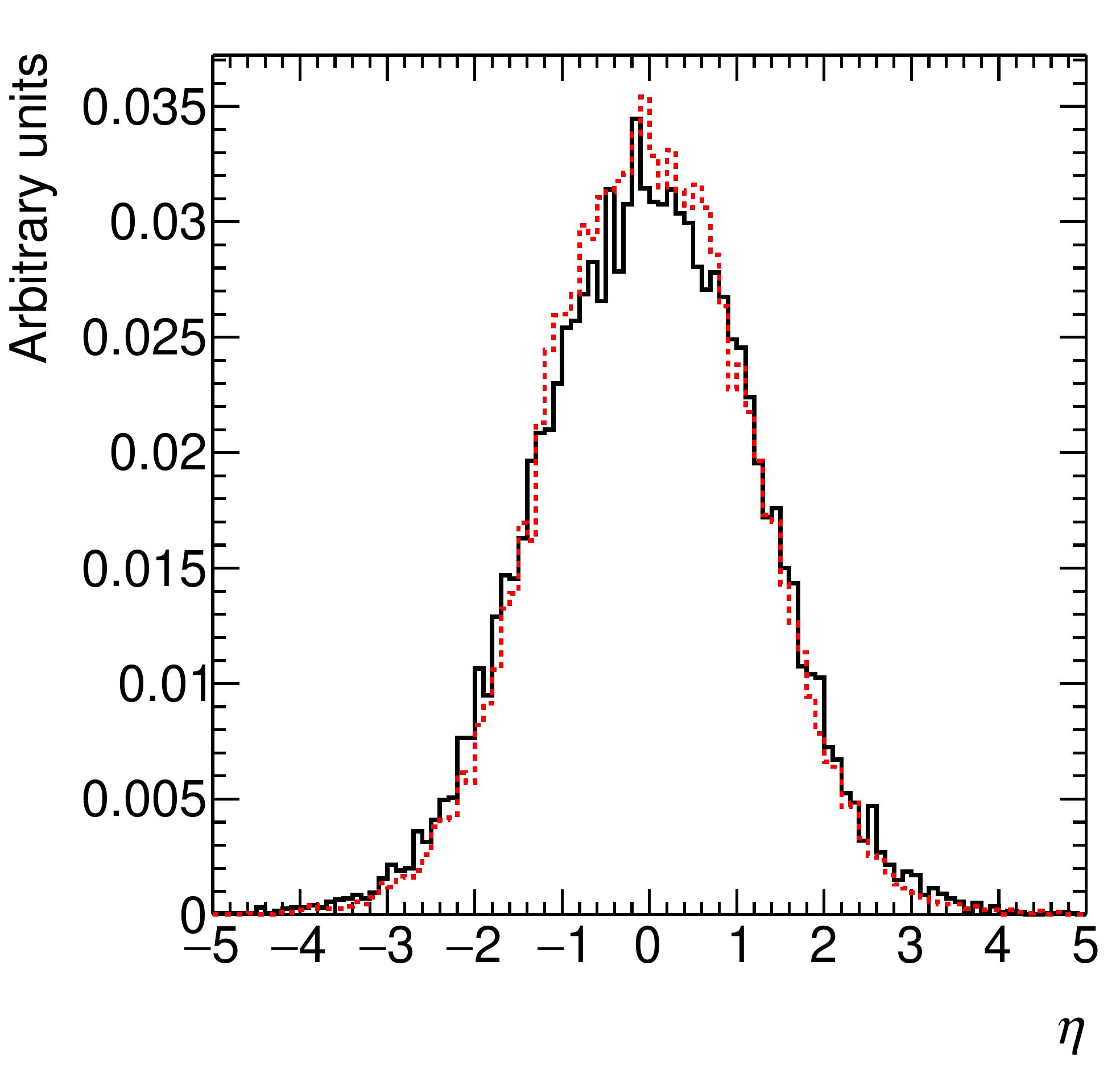}}\end{minipage}
\caption{\label{fig:kinematics_UPC}Without loss of generality, 1$g_\textrm{D}$ 
  monopoles were chosen for the transverse momentum~$p_\text{T}$ and
  pseudorapidity~$\eta$ of spin-0 and spin-\textonehalf\ HIPs with
  masses 40~GeV produced in lead-lead ultraperipheral collisions
  (UPC).}
\end{figure}

The kinematic distributions for ultraperipheral production of
1$g_\textrm{D}$ monopoles with mass 40~GeV are presented in
figure~\ref{fig:kinematics_UPC}. It should be noted that the
kinematics are independent of HIP charge. The transverse momentum
distribution is softer for spin-0 HIPs than for
spin-\textonehalf\ HIPs, as a result of the four-point vertex in the
spin-0 photon-fusion mechanism. The pseudorapidity distributions are
very similar in the two spin scenarios.

\begin{figure}[!h]
    \centering
    \begin{minipage}{0.45\textwidth}{
    \includegraphics[width=\textwidth,keepaspectratio=true]{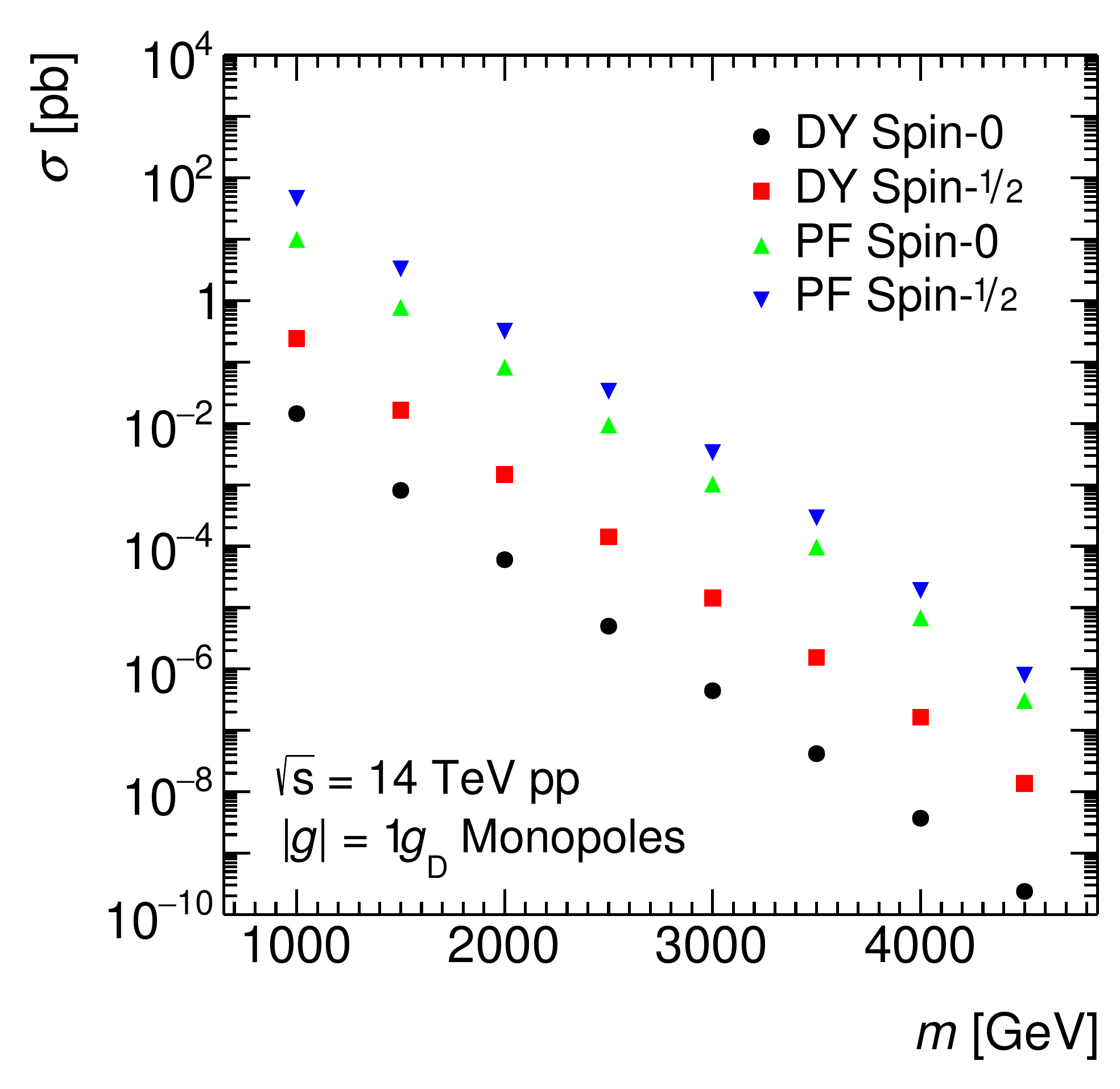}}\end{minipage}
\begin{minipage}{0.45\textwidth}{
    \includegraphics[width=\textwidth,keepaspectratio=true]{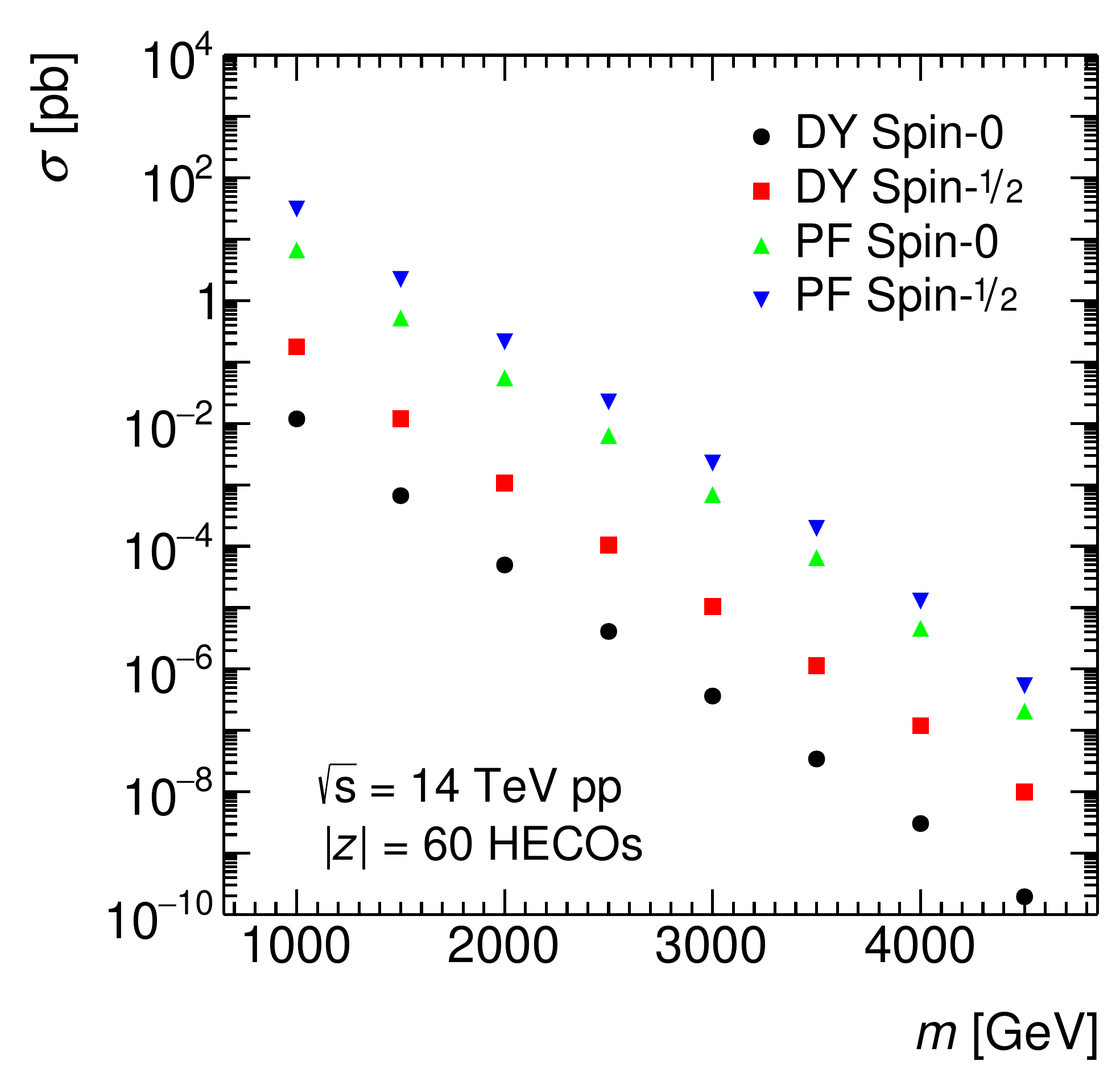}}\end{minipage}
    \caption{\label{fig:xsec_pp}Cross sections for spin-0 and
      spin-\textonehalf\ HIPs by Drell-Yan (DY) and photon fusion (PF) in
      proton-proton collisions.}
\end{figure}

Finally, we compare the cross sections for the Drell-Yan and
photon-fusion mechanisms in proton-proton collisions. The typical
masses considered in the LHC proton-proton collisions are on the order
of 1~TeV. As shown in figure~\ref{fig:xsec_pp}, the cross sections are
higher for photon fusion than for Drell Yan production. As was the
case in the ultraperipheral collisions, the cross sections for
spin-\textonehalf\ HIPs are larger than those for spin-0 HIPs.

In conclusion, we have studied the Drell-Yan and photon-fusion
mechanisms of scalar and fermionic HIP production in
$\sqrt{s_{\text{NN}}} = 5.5$~TeV heavy-ion ultraperipheral collisions
and $\sqrt{s} = 14$~TeV proton-proton collisions. The HIP kinematics
are significantly more limited in the ultraperipheral collisions,
hence, for comparison purposes their production in both collision
types is examined in a mass range that is accessible in
ultraperipheral collisions. In this mass range, their production in
lead-lead ultraperipheral collision is significantly enhanced due to
the large nuclear charge compared to the photon-fusion process in
proton-proton collisions. Although this difference plays no role at
the parton level on the matrix elements, it enhances the luminosity of
the effective photon collisions.  The spin-\textonehalf\ HECO
production has a non-trivial interference effect between the
$Z^0$-boson exchange and the photon exchange in the Drell-Yan
mechanism, giving rise to a slightly smaller cross section than the
exclusive photon-exchanged Drell-Yan mechanism. Nevertheless, the
photon-fusion cross section exceeds that of Drell-Yan production in
proton-proton collisions for the considered mass range as a
consequence of its squared dependence of the HIP-photon coupling
prevailing over the linear dependence of this coupling in the
Drell-Yan cross section.

\begin{acknowledgments}
We acknowledge Philippe Mermod's numerous contributions that led us to
pursue this work. We thank Arka Santra, Ameir Shaa Bin Akber Ali,
Stephanie Baines, Justin Kerr and Yury Smirnov for useful discussions
on the modelling of the photon-fusion mechanism in $pp$ collisions.
We thank Olivier Mattelaer for expert advice regarding implementing
PDFs into \textsc{MadGraph}5\_aMC@NLO. We acknowledge the support of
the Natural Sciences and Engineering Research Council of Canada
(NSERC).
\end{acknowledgments}

\bibliography{ms}

\end{document}